%% file: ms.tex
\documentclass[iop]{emulateapj}
\usepackage{amsmath,graphicx,longtable,multirow}    
\usepackage{wasysym}
\usepackage{xcolor}     
\usepackage{epsfig}
\usepackage{epstopdf}
\usepackage[]{natbib}

\def\msol{\hbox{$\rm\thinspace M_{\odot}\thinspace$}}

\def\eg{{\it e.g.\ }}

\newcommand{\be}{\begin{equation}}
\newcommand{\ba}{\begin{eqnarray}}
\newcommand{\ee}{\end{equation}}
\newcommand{\ea}{\end{eqnarray}}

\newcommand{\kms}{\ensuremath{\mathrm{km~s}^{-1}}}

\newcommand{\Mch}{\ensuremath{M_{\rm ch}}}

\newcommand{\Msun}{\ensuremath{M_\odot}}
\newcommand{\degree}{\ensuremath{^\circ}}
\newcommand{\unit}[1]{\ensuremath{\,\mathrm{#1}}}
\newcommand{\gcc}{\ensuremath{\unit{g\,cm^{-3}}}}
\newcommand{\cms}{\ensuremath{\unit{cm\,s^{-1}}}}
\newcommand{\el}[2]{$^{#2}\text{#1}$}

\input{nuclides.tex}

\shorttitle{Post-Merger Detonations}
\shortauthors{Raskin, Kasen, Schwab, Moll, \& Woosley 2013}

\begin{document}

\title{Type Ia Supernovae from Merging White Dwarfs II) Post-Merger Detonations}
\author{Cody Raskin\altaffilmark{1}, Daniel Kasen\altaffilmark{1,2}, Rainer Moll\altaffilmark{3}, Josiah Schwab\altaffilmark{2}, \& Stan Woosley\altaffilmark{3}}
\altaffiltext{1}{Nuclear Science Division, Lawrence Berkeley National Laboratory, Berkeley, CA, USA} 
\altaffiltext{2}{Departments of Physics and Astronomy, University of California, Berkeley, CA, USA}
\altaffiltext{3}{Departments of Physics and Astronomy, University of California, Santa Cruz, CA, USA}

\begin{abstract}

Merging carbon-oxygen (CO) white dwarfs are a promising progenitor
system for Type Ia supernovae (SN~Ia), but the underlying physics and
timing of the detonation are still debated.  If an explosion occurs
after the secondary star is fully disrupted, the exploding
primary will expand into a dense CO medium that may still have a
disk-like structure.  This interaction will decelerate and distort the
ejecta.  Here we carry out multi-dimensional simulations of ``tamped"
SN~Ia models, using both particle and grid-based codes to study the
merger and explosion dynamics, and a radiative transfer code to
calculate synthetic spectra and light curves.  We find that
post-merger explosions exhibit an hourglass-shaped asymmetry, leading
to strong variations in the light curves with viewing angle.  The two
most important factors affecting the outcome are the scale-height of
the disk, which depends sensitively on the binary mass ratio, and the
total \nickel[56] yield, which is governed by the central density of
the remnant core.  The synthetic broadband light curves rise and
decline very slowly, and the spectra generally look peculiar, with
weak features from intermediate mass elements but relatively strong
carbon absorption.  We also consider the effects of the viscous
evolution of the remnant, and show that a longer time delay between
merger and explosion probably leads to larger \nickel[56] yields and
more symmetrical remnants.  We discuss the relevance of this class of
aspherical ``tamped" SN~Ia for explaining the class of
``super-Chandrasekhar'' SN~Ia.

\end{abstract}

\section{Introduction}
There are currently two broad classes of models for SN Ia.  In the
first so-called ``single degenerate'' scenario, a CO white dwarf (WD)
accretes material from a main sequence or post-main sequence companion
\citep{Whelan1973,Nomoto1982a}. As the WD grows and approaches the
Chandraskehar mass (\Mch), its central density increases, and a
subsonic deflagration burning front is ignited near the center, likely
later transitioning to a detonation \citep[see][and references
  therein]{Hillebrandt2000}. Alternatively, a detonation may occur in
a surface layer of accreted helium before the WD has reached \Mch,
setting up a secondary detonation of the WD core and a subsequent
SN~Ia \citep{Woosley1994, Livne_1995,
  Fink2007,Fink2010,Ruiter2011,Sim2012,Moll2013}.

The second, ``double degenerate'' scenario invokes the dynamical merger of two WDs in a close binary \citep{Iben1984,Webbink1984,Benz1990,Yoon2007,Pakmor2010}. Since WDs are supported by degeneracy pressure, their inverse mass-radius scaling relationship often leads to catastrophic mass loss once one of the pair of WDs overflows its Roche lobe. \cite{Marsh2004} have demonstrated that for a large parameter space of possible WD binary masses, mass transfer is  unstable and leads to the complete disruption of the less massive companion star. 

When such a merger occurs, there are several possible outcomes. On the
one hand, the violent nature of the mass transfer itself may initiate a
prompt detonation in the primary WD as material from the companion
accretes onto its surface on a dynamical timescale. This outcome of
such ``peri-merger" detonations has been considered by
\cite{Pakmor2010,Pakmor2011} and \cite{Guillochon2010} and is the
subject of a companion paper \citep{Moll+2013}. In such a scenario,
the companion WD has only begun to be tidally disrupted, and remains
largely intact at the time of detonation.

On the other hand, the companion might be fully disrupted before a
violent detonation takes place, forming a disk around the primary
WD. Shear heating and compression of the differentially rotating
remnant may possibly set off a detonation in the primary
\citep{vanKerkwijk2010, Shen2012}.  If neither of these outcomes
result, the disk will evolve viscously to a more spherical geometry
\citep{Schwab2012}, and as the degenerate core is compressed on a
longer time scale, it may either explode or collapse to form a neutron
star \citep{Saio1985,Yoon2007,Shen2012}. \cite{RaskinKasen2013} have
considered how these outcomes might be distinguished by looking for
the observational signatures of material that has been dynamically
ejected during the merger process.

WD cores that explode inside the envelope of a disrupted secondary star would resemble the ``tamped" SN~Ia models considered originally by \cite{Khokhlov1993} and \cite{Hoeflich1996}.  In those parameterized 1D models, a near-\Mch\ degenerate CO WD was assumed to explode into a spherical medium of CO.  The interaction of the SN ejecta with the CO medium decelerated the ejecta, piling up material into a dense shell.  Because the  surrounding medium was  assumed to be at relatively small radii ($r \lesssim 10^{10}$~cm) any direct emission from the interaction would be difficult to detect. The restructuring of the SN ejecta has an impact on the light curves and spectra, however, and a similar shell structure has been invoked to explain the 1--2 week velocity plateau seen in some SN~Ia \citep{Quimby2007}.   \cite{Fryer2010} also considered the case where the CO medium was more extended and ongoing interaction influenced the brightness of the light curve.

The tamped SN~Ia models from WD merger have recently gained more interest as a possible explanation for the  class of so called ``super-\Mch'' SN~Ia \citep{Howell2006, Hicken2007,Scalzo2010,Taubenberger2011,Silverman2011}.   These rare events are nearly twice as bright as the most common SN~Ia, have broad light curves, and often show strong carbon lines in their spectra.  In addition, the ejecta velocities, as measured from the Doppler shift of the SiII line, are often $\sim 20\%$ lower than those of normal SN~Ia. In an empirical analysis  of SN~2007if,  \cite{Scalzo2010} inferred an ejecta masses of $M = 2.4 \pm 0.2~\Msun$ and a \nickel[56] mass of $M_{\rm ni} = 1.6 \pm 0.1~\Msun$.  These estimates explicitly assumed  a shell-like ejecta structure  such as that in a tamped SN~Ia model.  Estimates for SN~2009dc were in a similar range, $M \gtrsim 2.0~\Msun$ and $M_{\rm ni} \approx 1.4-1.8~\Msun$ \citep{Taubenberger2011,Silverman2011}.  To explain the strong observed carbon lines, a relatively large mass of unburned CO is apparently required \citep{Hachinger_2012}.  

These inferred properties of the super-\Mch\ SN are difficult to
reconcile with CO WD explosions, either in the single or double
degenerate scenario \citep{Maeda2009}. In rare cases, the merger of
two massive WDs might have a total ejected mass exceeding 2.0~\Msun,
but the production of $\gtrsim 1.5~\Msun$ of \nickel[56] would still
be difficult.  In simulations where a detonation is assumed to occur
early in the merger process, \nickel[56] is synthesized only in the
primary WD, because the tidal distortion of the secondary WD usually
reduces its density below the threshold for burning to nuclear
statistical equilibrium.  This limits the total \nickel[56] produced
in peri-merger detonations to be near or less than the mass of the
more massive WD.  If an explosion occurs post-merger, however, the WD
primary will be compressed by the coalescence with the secondary, and
the central density of the merged remnant is likely to be higher, 
allowing for a greater \nickel[56] production.

One limitation of most empirical analyses of the super-\Mch\ SN has
been the assumption of spherical symmetry.  If the SN ejecta is
asymmetric, the resulting emission will be anisotropic and, depending
on the viewing angle, the \nickel[56] mass required to produce the
observed peak luminosity could be reduced \citep{Hillebrandt_2007,
  Kasen2007}.  Simulations of peri-merger detonations find that the
ejecta is highly asymmetric due to the interaction of the exploded
primary with the nearly intact secondary star.  For cases with massive
($\sim 1.2~\Msun$) CO WD primaries, the light curves can approach the
observed luminosities of the super-\Mch\ events from some viewing
angles \citep{Moll+2013}.  If, however, the detonation occurs at a
later phase, the secondary will have been tidally disrupted into a
disk.  The resulting asymmetry and orientation effects will be
distinctive and have not been examined before.

In this paper, we calculate the observational properties of
multi-dimensional, tamped SN~Ia from WD mergers, and consider, in
particular, their relevance to super-\Mch\ SN.  A combination of
lagrangian and grid-based hydrodynamical tools is used to model the
merger dynamics and explosion, and a separate radiation transport code
is employed to generate angle-dependent light curves and spectra.  In
a companion paper \citep{Moll+2013}, the results of simulations of
early-time detonations in these systems was described.  Here we focus
on the later stages of the WD merger process where only a single
degenerate core remains, enshrouded by either a disk or a more
spherical distribution of matter \citep{Schwab2012}.  The synthesis of
\nickel[56] is determined and the observable properties calculated.
Section 2 describes the numerical hydrodynamics methods, initial
conditions, and merger and detonation results. Section 3 reports the
results of the radiation-transport calculations and Section 4
discusses our results.

\section{Merger Simulations \& Detonations}

Smoothed particle hydrodynamics (SPH) codes are especially useful for
simulating compact object mergers since they conserve angular momentum
explicitly. This is crucial for properly modeling the resultant merger
remnant configuration. Following the procedure of \cite{Raskin2012},
we simulate several pairs of WD binaries whose masses are given in
Table \ref{table:grid}, using \textsc{snsph} \citep{Fryer2006}. We use
the Helmholtz free-energy equation of state \citep[EOS:][]{TimmesArnett1999,Timmes2000} for its applicability to a
range of states from ideal gas to degenerate electron pressure
support, and including photon pressure support.

\begin{table}[ht]
\caption{Simulated binary mass pairs and the $q$-parameter. All masses are solar.}
\centering
\begin{tabular}{c | c l | c c | c}
\hline\hline
\# & $m_1$ & $m_2$ & $q$ $(m_2/m_1)$ \\
\hline
0p9-0p6& 	0.96 & 	0.64 &	0.67\\
0p9-0p8& 	0.96 & 	0.81 & 	0.84\\
1p0-0p6& 	1.06 & 	0.64 & 	0.60\\
1p0-1p0& 	1.06 & 	1.06 & 	1.00\\
1p2-0p6& 	1.20 & 	0.64 & 	0.53\\
1p2-1p0& 	1.20 & 	1.06 & 	0.88\\
1p0-0p4&	1.06 & 	0.40 He &	0.38\\
\hline
\end{tabular}
\label{table:grid}
\end{table}

Each of our stars consists of $4\times10^{5}$ particles of approximately equal mass, arranged using weighted Voronoi tessellations \citep[\textsc{wvt}:][]{WVT}. Such a configuration, sometimes called a Dirichlet tessellation, maximizes compactness for particles of varying size (smoothing length) and greatly reduces the time to relaxation over more glass-like arrangements. Greater compactness also prevents entropy reversals that can sometimes result from high entropy particles \textit{slipping through} cracks in more uniform particle arrangements, and does so without the need for an unsuitably large artificial viscosity. Since interactions between particles in SPH codes are typically dominated by the most massive particle in the interaction, constraining our particles to roughly equal masses ensures that the interaction of the accretion flow from the secondary star with the surface material of the primary is well resolved and robust for our purposes. Microphysical processes such as those responsible for detonations cannot be resolved at these scales, however. 
All stars are initially isothermal with a temperature of $\approx5\times10^4$ K with random temperature fluctuations of order $\sim10^4$ K. Each star has a constant composition of 50\% \carbon[12] and 50\% \oxygen[16] (CO) by mass fraction, with the exception of the 0.40\msol star in simulation 1p0-0p4, which is initialized with 100\% \helium[4]. Our simulations employ the $\alpha$-chain 13-isotope nuclear burning network, \textsc{aprox13} \citep{Timmes1999}, in order to capture any nuclear burning that might occur during the merger phase. This is important since nuclear energy generation plays a part in setting the scale height of the remnant disk.

\begin{figure}[ht]
\centering
\includegraphics[width=0.36\textwidth]{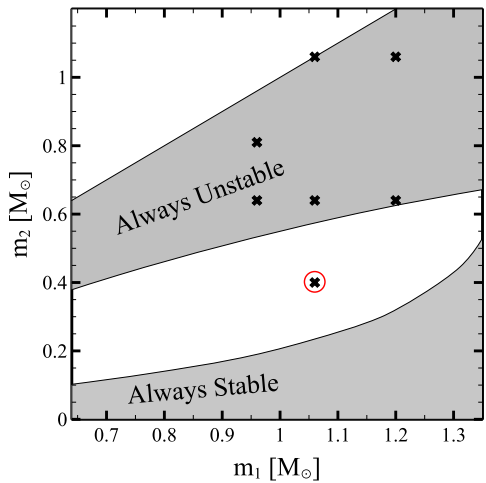}
\caption{Stable and unstable mass transfer regimes for binary WD mergers, reproduced from \cite{Marsh2004} with our simulations indicated by crosses. The circled cross, simulation 1p0-0p4, lies in the region of parameter space between always unstable and always stable.}
\label{fig:stability}
\end{figure}

\cite{Marsh2004} studied the dissipative forces involved in
binary mass transfer for WDs and gave analytical estimates of
the parameter space for stable and unstable mass transfer. In this
case, unstable mass transfer refers to a binary system wherein, once
mass transfer begins, the secondary star continues to lose mass at an
accelerated rate until its complete disruption on a dynamical
timescale. As seen in Figure \ref{fig:stability}, which reproduces the
analytical estimates of the stable and unstable parameter space, all
of our simulations except for simulation 1p0-0p4 lie well within the
regime of unstable mass transfer. 1p0-0p4 lies in the region between
stability and instability, and as will be discussed, this simulation
also exhibits unstable mass transfer.

\begin{figure}[ht]
\centering
\includegraphics[width=0.40\textwidth]{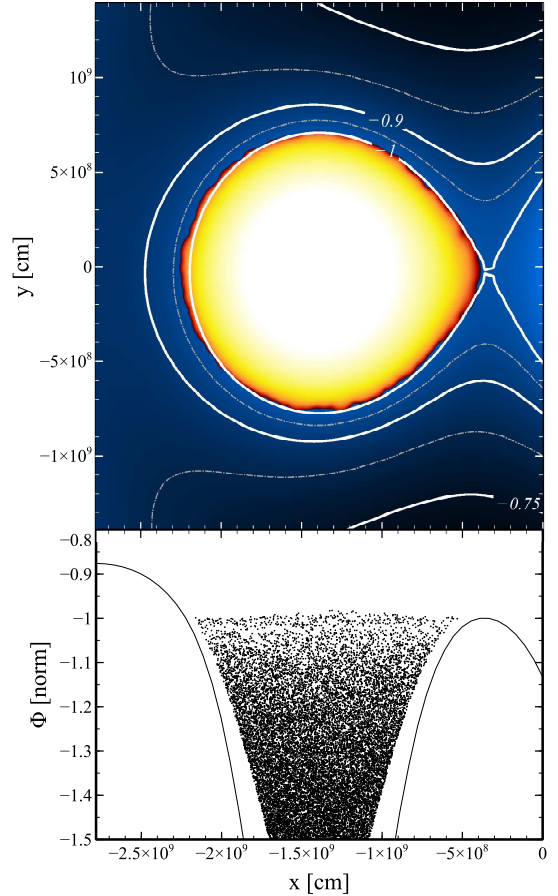}
\caption{A rotating-frame, gravitational potential map for the initial conditions of simulation 1p0-0p6. In the top frame, a density slice of the 0.64\msol star is overlaid with potential contours, with the first contour at the potential energy of the L1 point, where the potential has been normalized to unity. In the bottom frame, a representative subsample of particles is plotted against a trace of the system's potential through the \textit{x}-axis. As is evident, the star has filled its Roche lobe, but no single particle has crossed the potential barrier between the two stars.}
\label{fig:pot}
\end{figure}

Additionally, \cite{Dan2011} have shown that initial conditions which place the WDs too near each other often result in unrealistic accretion rates and can distort the final remnant configuration. Too vigorous a flow can result in hotter disk temperatures, and thus a larger scale height. As in \cite{Raskin2012}, we ensure robust initial conditions by first relaxing the WD pair in their combined gravitational potential. We then move the stars nearer each other gradually until particles residing in the tidal bulge of the less massive WD are just below the rotating-frame potential maximum between the two stars. In this way, the merger simulation begins with synchronized WDs that are already tidally distorted (and not ringing), and with the less massive WD just beginning to overflow its Roche lobe. Figure \ref{fig:pot} demonstrates this for the initial conditions of simulation 1p0-0p6, wherein the 0.64\msol WD has filled its Roche lobe, but no single particle has yet traversed the potential barrier between the two stars. In this case, $\Phi=\Phi_g - 0.5\Omega^2r^2$ and is normalized to $-1$ at L1.

Initializing our simulations this way ensures that the resultant
remnant disks after accretion will be concentrated along the equator
and not overly spherical as a result of excessive heating during an
unphysically vigorous mass-exchange. As will be seen in later
sections, the scale height and equatorial concentration of the disk
play a crucial role determining the angle-dependent spectra of these
models.

\subsection{Remnant Configurations}

\begin{figure}[ht]
\centering
\includegraphics[width=0.40\textwidth]{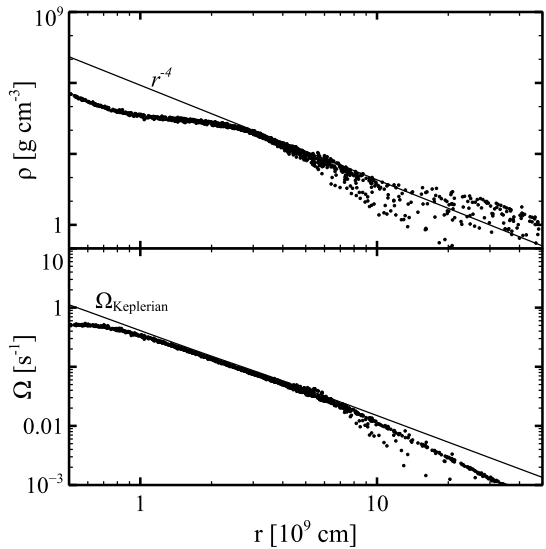}
\caption{Disk profiles for simulation 1p0-0p6. The upper panel displays the density profile of the disk, roughly reproducing the $r^{-4}$ profile of \cite{Fryer2010}, while the lower panel demonstrates the sub-Keplerian rotation rate of the disk.}
\label{fig:1p0-0p6disk}
\end{figure}

For all the merger simulations, including 1p0-0p4, there is an
extended accretion phase that persists for several orbits. Some
material is lost from the companion and becomes unbound from the
system during this phase through the L2 point ($M_{\rm
  ej}\sim10^{-3}$\msol), as described in
\cite{RaskinKasen2013}. Eventually, the companion star is completely
disrupted and forms a hot accretion disk around the primary. For
simulation 1p0-1p0, small asymmetries between the two stars were
sufficient to unbind one of the stars rather than both. This mechanism
is described in more detail in \cite{Raskin2012}. The result of each
of these simulations is a mostly degenerate core surrounded by a
thermalized disk of \carbon[12] and \oxygen[16]. The exception is
simulation 1p0-0p4, where the disk is comprised almost entirely of
\helium[4] with faint traces of \carbon[12] and \oxygen[16] lost from
the surface of the primary and as a result of some very mild burning
during the merger. All of the material that rotates as a solid body is
assumed to be the core, with the rest of the material comprising a hot
disk envelope. The majority of the disk material follows a roughly
$r^{-4}$ profile as was found in \cite{Fryer2010}. Figure
\ref{fig:1p0-0p6disk} demonstrates that much of the disk is
sub-Keplerian as it is partially thermally supported.

In general, the scale heights of the disks are correlated with the
$q$-values of the binary systems, with smaller $q$-values resulting in
larger scale heights. For scale heights of $H\approx2-4\times10^8$
cm, listed in Table \ref{table:heights}, we expect the viscous
evolution of these disks to behave like that of thick disks, with
evolution times $\sim10$ hours \citep{vanKerkwijk2010}. It is notable
that simulation 1p0-0p4 features a remnant disk with a scale height of
$H\approx10^9$ cm. The disruption of the companion in this simulation
was much more catastrophic than in the CO-CO mergers since small changes
in mass lead to larger increases in the radius of an already reduced
mass companion. In addition, helium burning liberates
considerably more energy in this simulation, thermalizing the disk to
a greater extent than in the CO-CO mergers. This large discrepancy in
the scale heights between the CO-CO mergers and simulation 1p0-0p4 has
an important impact on the variability of the spectra at different
viewing angles of the SN produced inside these disks. This effect will
be discussed further in \S3.

\begin{table}[ht]
\caption{Scale heights of the remnant disks of each of our merger simulations.}
\centering
\begin{tabular}{c | c c}
\hline\hline
\# & $q$ & $H$ [$10^8$ cm]\\
\hline
0p9-0p6& 	0.67 & 4.37\\
0p9-0p8& 	0.84 & 2.18\\
1p0-0p6& 	0.60 & 4.21\\
1p0-1p0& 	1.00 & 2.50\\
1p2-0p6& 	0.53 & 4.05\\
1p2-1p0& 	0.88 & 2.58\\
1p0-0p4&	0.38 & 9.59\\
\hline
\end{tabular}
\label{table:heights}
\end{table}

For the most part, little nuclear burning takes place during the
merger process. Though the SPH simulations of many of the mergers do
reach detonation conditions at the surface of the primary as
prescribed by \eg \cite{Seitenzahl2009}, these simulations,
nevertheless, do not detonate self-consistently. There are a variety
of possible explanations for this, some of which are
code-dependent. \cite{Guillochon2010} have found that in similar
models with highly resolved accretion flows of pure \helium[4],
self-initiating surface detonations can sometimes result. However,
hydrodynamical simulations of the scale presented in this paper cannot
accurately resolve the detonation conditions inside the accretion
flow. Moreover, it remains an open question as to how WD mergers or
their remnants ignite in nature, and since we aim to quantify the
effects of the remnant disk (or viscously evolved envelopes) on the
light curves and spectra of SN~Ia embedded within them, for this paper,
we chose not to insert artificial detonations at these early stages
in the merger simulations in order to explore the consequences.

\subsection{Explosion Dynamics with \textsc{snsph}}

Typically, conditions that lead to a self-consistent detonation in an
SPH calculation are stricter than those of grid codes, and surface
detonations are notoriously difficult to ignite. We do not expect the
question of central ignition or surface ignition to drastically affect
our conclusions of how the SN shock interacts with the disk, since
this detail mainly concerns the production of \nickel[56] and not the
detonation geometry, and as will be discussed, even the \nickel[56]
yield is not drastically altered. For these reasons, we artificially
detonate our merger remnants via an instantaneous conversion of the
unburned remnant core to burned ash.

\begin{figure}[ht]
\centering
\includegraphics[width=0.40\textwidth]{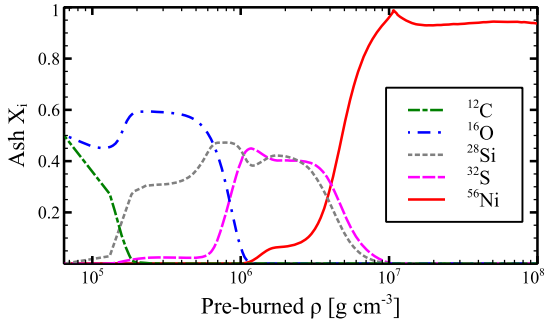}
\caption{Ash composition as a function of pre-burned density for a 1.2\msol WD detonation as calculated by \textsc{kepler}. Only the most abundant isotopes are plotted.}
\label{fig:abund}
\end{figure}

During the merger phase of the calculation, we used \textsc{aprox13} to capture any nucleosynthesis that may occur from sub-critical nuclear burning at the shock front of the accretion stream with the primary. Once a quasi-equilibrium was reached, the hydrodynamics calculation was taken offline, and all material
 at densities $>10^6$ g
cm$^{-3}$ was instantaneously burned. In order to determine the composition of the incinerated material, a
series of 1D white dwarf detonations of various masses was calculated
offline using the \textsc{kepler} 1D implicit hydrodynamics package
\citep{Weaver1978,Woosley2002}.  The results were used to create a
lookup table that was employed for the instantaneous conversion phase
in \textsc{snsph}. The network in these \textsc{kepler} calculations was
adaptive in time \citep{Rauscher2002}, and for these studies, included between 380 to 564 isotopes coupled
directly to the nuclear energy generation at each time
step. The lookup table produced this way provides a post-detonation
ash abundance vector with solar metallicity pollutants for a range of
pre-shock densities of sub-Chandrasekhar WDs
\citep{WoosleyKasen2011}. This ash composition strongly depends on the
core density, and since the variation in the total \nickel[56] yield
between any two of our pre-computed models is less than 5\% by mass,
we interpolated between the results of \textsc{kepler} for various
initial core masses to approximate abundances for a range of central
densities. Figure \ref{fig:abund} shows a sample ash composition for a
range of pre-shock densities from a \textsc{kepler} calculation of a
1.2\msol WD detonation.

We deposit the nuclear energy release (minus neutrino
losses) as thermal energy.
This unbinds the star and leads to explosion shock that propagates through the disk.
We simulate this shock 
and expansion phase using normal hydrodynamics, again coupled to \textsc{aprox13}. 
The approach is based on the fact that the nucleosynthesis
in a supersonic detonation is primarily a function of the density of
the fuel.  However, our instantaneous deposition of energy neglects
the expansion of shocked material that will occur while the detonation
is propagating through the remnant star. In a later section,
we compare the results obtained this way with those computed
self-consistently with a grid code, and show that the results are very
similar.

Using \textsc{aprox13} for the hydrodynamical portion of the detonation simulation after the instantaneous conversion of the core allows us to capture $\alpha$-chain nuclear burning in the portions of the core below $>10^6$ g cm$^{-3}$ and in the disk as the SN shock overtakes it. Specifically, this includes triple-$\alpha$ burning in the helium disk of simulation 1p0-0p4. The hydrodynamical time step in all cases is limited to the smaller of either the Courant condition or the burning timescale that results in at most a 30\% change in internal energy. Unstable isotopes, such as \iron[52], which has a half-life of $\approx8$ hours, are included in the calculation of the nuclear energy released during burning, but are decayed to their stable daughter isotopes during the expansion phase. The radiation energy from these decay processes is expected to be negligible compared to that of \nickel[56], which has a half-life of roughly 6 days, and so we do not track the energy deposition from the decay of these short-lived nuclei. 

\subsubsection{\textsc{snsph} Detonations}

\begin{figure*}[ht]
\centering
\includegraphics[width=0.80\textwidth]{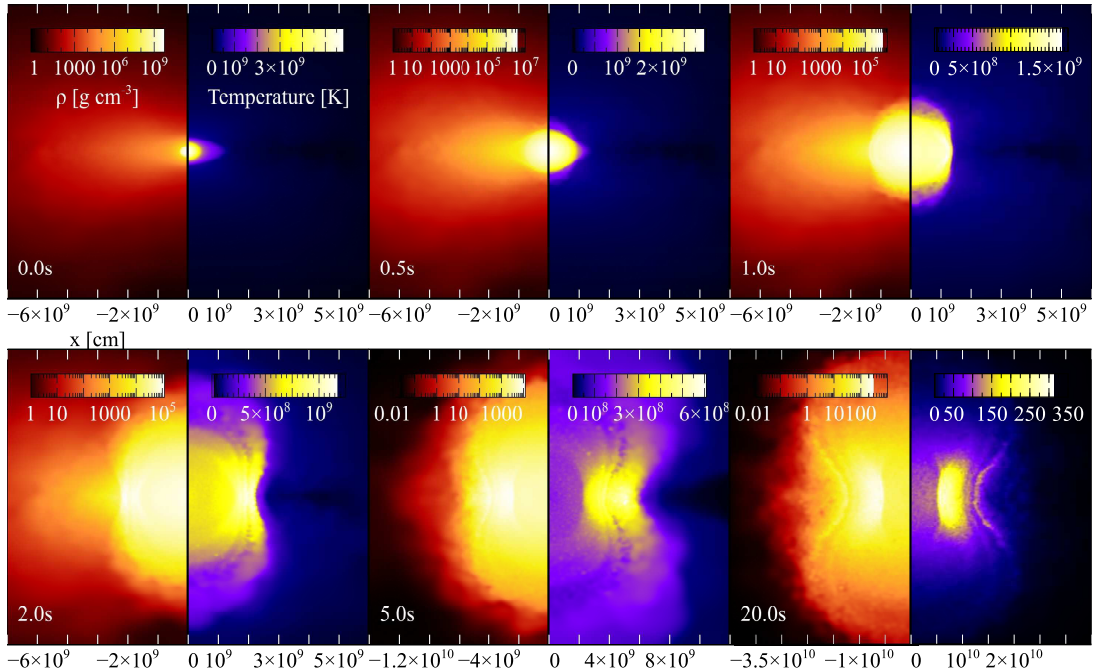}
\caption{Time evolution of the shock geometry of simulation 0p9-0p8 as the detonation shock interacts with the remnant disk depicted with slices in the $x$-$z$ plane (each time-frame is square).}
\label{fig:diskevo}
\end{figure*}

Figure \ref{fig:diskevo} illustrates the time evolution of the shock as it encounters the remnant disk at the equator. While it expands relatively freely at the poles, the high density disk material acts to slow the SN ejecta along the equator. The result is an hourglass shape and a strong global asymmetry that will influence the light curves and spectra. 

\begin{figure}[ht]
\centering
\includegraphics[width=0.40\textwidth]{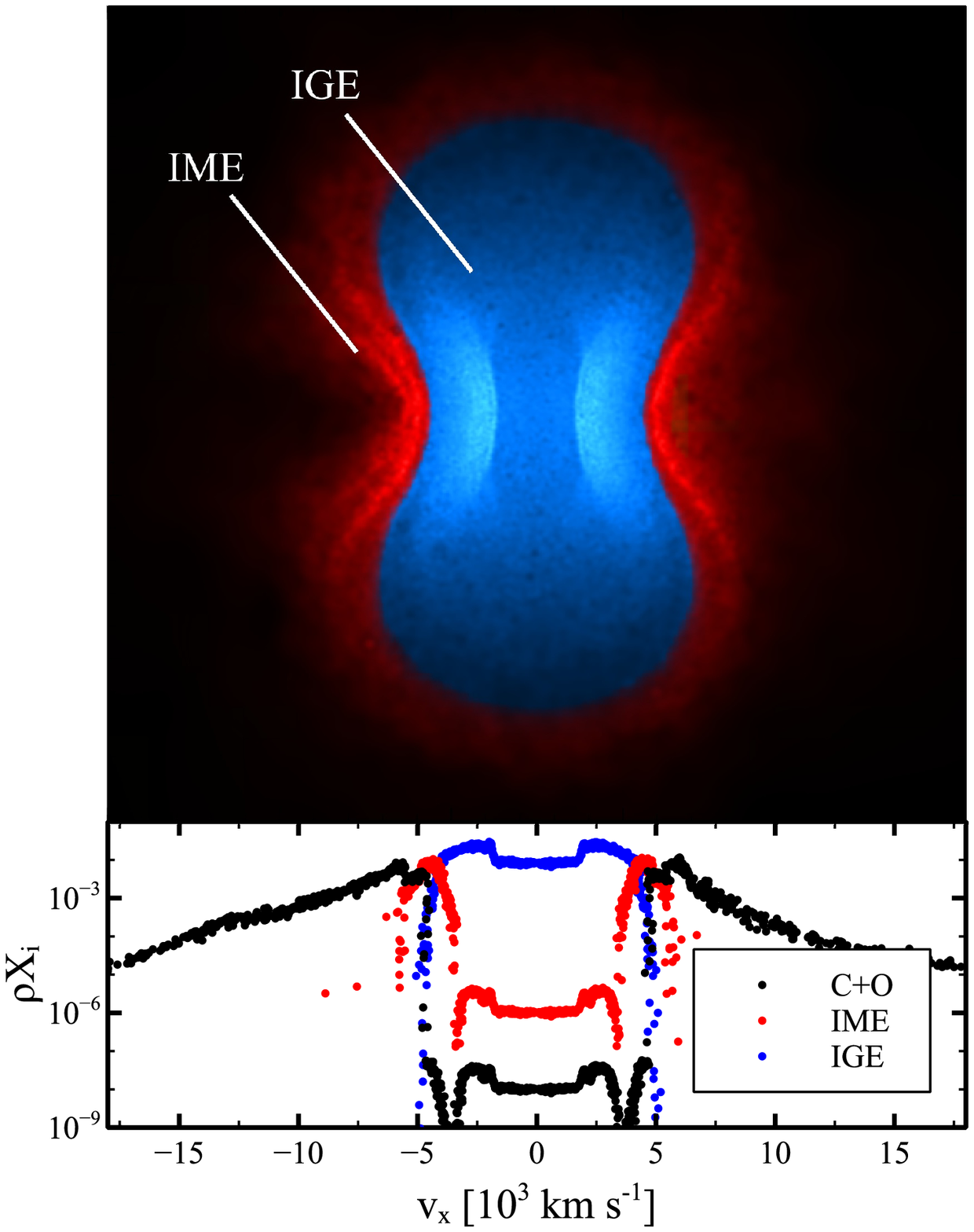}
\caption{Isotopic stratification in the supernova remnant of simulation 0p9-0p8 during the homologous expansion phase. The blue region indicates the highest concentrations of iron-group elements, while the red region demarcates the intermediate-mass element distribution. The brightest areas in each region represent the highest concentrations by mass-density. The rest of the domain in this image is almost entirely carbon+oxygen, as indicated in the lower panel which plots density (scaled by mass-fraction of each of the three major isotope groups) as a function of velocity.}
\label{fig:isotopes}
\end{figure}

The isotopic stratification during homology is illustrated in Figure \ref{fig:isotopes}. As expected, the iron-group elements (IGE) are concentrated in the central regions of the SN ejecta, while intermediate-mass elements (IME) and CO make up the outer layers respectively. In the plane of the disk, both the IGE and IME are slowed due to the interaction of the SN shock with the remnant disk. We will discuss the details of the light curves and spectra in \S3.

In Table \ref{table:sphyields}, we compare the nucleosyntheic yields for our \textsc{snsph} simulations. Since the \nickel[56] yield depends on the central density of the remnant core, the final product varies with both the primary mass and the total mass. 

\begin{table*}[ht]
\caption{Isotope yields and kinetic energy for our \textsc{snsph} simulations.}
\centering
\begin{tabular}{r | r r r r r r r}
\hline\hline
&0p9-0p6&0p9-0p8&1p0-0p6&1p0-1p0&1p2-0p6&1p2-1p0&1p0-0p4\\
\hline
\helium[4]&$7.54\times10^{-3}$&$6.48\times10^{-3}$&$1.18\times10^{-2}$&$1.20\times10^{-2}$&$1.44\times10^{-2}$&$1.50\times10^{-2}$&0.390\\
\carbon[12]&0.228&0.229&0.251&0.226&0.261&0.234&$2.09\times10^{-2}$\\
\oxygen[16]&0.315&0.384&0.318&0.452&0.291&0.376&$3.80\times10^{-2}$\\
\neon[20]&$3.41\times10^{-3}$&$1.23\times10^{-2}$&$3.12\times10^{-3}$&$1.46\times10^{-2}$&$1.40\times10^{-3}$&$1.14\times10^{-2}$&$1.13\times10^{-3}$\\
\magnesium[24]&$1.39\times10^{-2}$&$3.04\times10^{-2}$&$1.08\times10^{-2}$&$4.10\times10^{-2}$&$3.75\times10^{-3}$&$2.21\times10^{-2}$&$4.91\times10^{-3}$\\
\silicon[28]&0.155&0.215&0.107&0.257&0.0546&0.168&0.120\\
\el{S}{32}&0.121&0.152&0.0811&0.154&0.0414&0.1033&0.115\\
\el{Ar}{36}&$2.12\times10^{-2}$&$2.60\times10^{-2}$&$1.48\times10^{-2}$&$2.54\times10^{-2}$&$7.34\times10^{-3}$&$1.78\times10^{-2}$&$2.08\times10^{-2}$\\
\el{Ca}{40}&$1.90\times10^{-2}$&$2.27\times10^{-2}$&$1.36\times10^{-2}$&$2.11\times10^{-2}$&$6.27\times10^{-3}$&$1.49\times10^{-2}$&$1.88\times10^{-2}$\\
\el{Ti}{44}&$4.90\times10^{-4}$&$5.49\times10^{-4}$&$4.02\times10^{-4}$&$5.01\times10^{-4}$&$1.94\times10^{-4}$&$3.94\times10^{-4}$&$4.95\times10^{-4}$\\
\el{Cr}{48}&$9.75\times10^{-3}$&$1.09\times10^{-2}$&$7.63\times10^{-3}$&$8.92\times10^{-3}$&$2.81\times10^{-3}$&$6.08\times10^{-3}$&$9.72\times10^{-3}$\\
\el{Fe}{54}&$1.96\times10^{-2}$&$1.73\times10^{-2}$&$3.01\times10^{-2}$&$3.05\times10^{-2}$&$6.10\times10^{-2}$&$6.66\times10^{-2}$&$2.06\times10^{-2}$\\
\el{Ni}{56}&0.687&0.664&0.856&0.886&1.09&1.23&0.704\\
\hline
KE [erg]&$1.29\times10^{51}$&$1.35\times10^{51}$&$1.38\times10^{51}$&$1.65\times10^{51}$&$1.54\times10^{51}$&$1.82\times10^{51}$&$1.27\times10^{51}$\\
$28<A<40$&0.316&0.416&0.216&0.457&0.110&0.304&0.275\\
$A\geq44$&0.716&0.692&0.894&0.926&1.16&1.30&0.734\\\hline
\end{tabular}
\label{table:sphyields}
\end{table*}

\subsection{Explosion Dynamics with \textsc{castro}}
In order to ensure that the shock geometry and the details of the shock interaction with the disk are not spurious or method-dependent, we also employ an Eulerian hydrodynamics code, \textsc{castro} \citep{Castro,Zhang2011} for simulations 0p9-0p8 and 1p0-0p6 (henceforth referred to as 0p9-0p8c and 1p0-0p6c, respectively, when intended to refer to the \textsc{castro} counterparts to the \textsc{snsph} simulations.). Since \textsc{castro} simulations are computationally expensive as compared to \textsc{snsph} simulations, we restrict our exploration to these two comparators. The equation of state remains the Helmholtz free-energy EOS, and the nuclear energy calculations are restricted to a 199-isotope lookup table. The final remnant stage of simulation 0p9-0p8 was interpolated onto a 3-level static mesh. Each level consists of a cube centered on the point of highest density.  The domain sizes are $(12\times10^8\unit{cm})^3$, $(48\times10^8\unit{cm})^3$ and $(192\times10^8\unit{cm})^3$.  The grid is resolved by $320^3$ zones at each level in the 1p0-0p6c model, and $256^3$ zones at each level in the 0p9-0p8c model.  The innermost level was dropped when the detonation reached its boundaries. Once the ejecta neared the boundary of the largest domain, we mapped the data into a new domain twice as large, dropping the innermost level of refinement.  This step was repeated until the ejecta neared the boundaries of a domain of $(3.07\times10^{11}\unit{cm})^3$, at which time the expansion is largely homologous, and the internal energy is only a small fraction ($< 1.5\%$) of the kinetic energy.

Unlike the \textsc{snsph} simulations used for calculating the merging process, \textsc{castro} is unable to handle vacua.  The density of the ambient medium was set to $10^{-1}\gcc$ at the beginning, and lowered by factors of 10 down to $10^{-4}\gcc$ during remaps into bigger domains (to fill the volume not covered by the old domain).  The total mass of the ambient medium filling empty regions is always smaller than $1.9 \times 10^{-3} \Msun$.

To further increase the numerical stability during the shock breakout and to contain detrimental effects on the time step, we employed a velocity cap of $3\times10^9\cms\approx0.1c$. A tally of the total kinetic energy subtracted this way is smaller than $1.5\times10^{49}\unit{erg}$ at the end of the simulation, which is insignificant compared to the remaining kinetic energy.

In Table \ref{table:codecompare}, we compare the two major hydrodynamics codes used in this paper for simulating detonations, \textsc{castro} and \textsc{snsph}. As \textsc{snsph} is a particle code, it does not have a lower limit on the size of its interpolants - particles have a fixed mass and their sizes are determined by their densities at any time step. Typical particle sizes in high density regions are roughly ($\sim10^6$ cm)$^3$, but can vary considerably. \textsc{snsph} also does not have a fixed domain size - simulations can grow without bound as the interpolants are lagrangian.

\begin{table*}[ht]
\caption{A comparison of the hydrodynamics methods, \textsc{snsph} and \textsc{castro}, used to simulate detonations.}
\centering
\begin{tabular}{l | c c}
\hline\hline
 & \textsc{snsph} & \textsc{castro}\\
\hline
Initial Condition & Binary System & Post-Merger Remnant\\
Algorithm & 3D SPH & 3D Eulerian Mesh\\
Interpolants & $4\times10^5$ particles & $3\times$($320^3$/$256^3$) cells\\
Smallest Interpolant & ($\sim10^6$ cm)$^3$& ($4.69\times10^6$ cm)$^3$\\
Domain Size & -- & ($3.07\times10^{11}$ cm)$^3$\\
EOS & Helmholtz & Helmholtz\\
Reaction Network & \textsc{aprox13} + lookup table & lookup table\\
Ignition Location & Global ($\rho>10^6$ g cm$^{-3}$) & Edge-lit\\
\hline
\end{tabular}
\label{table:codecompare}
\end{table*}

\subsubsection{\textsc{castro} Detonations}

\begin{figure}[ht]
\centering
\includegraphics[width=0.4\textwidth]{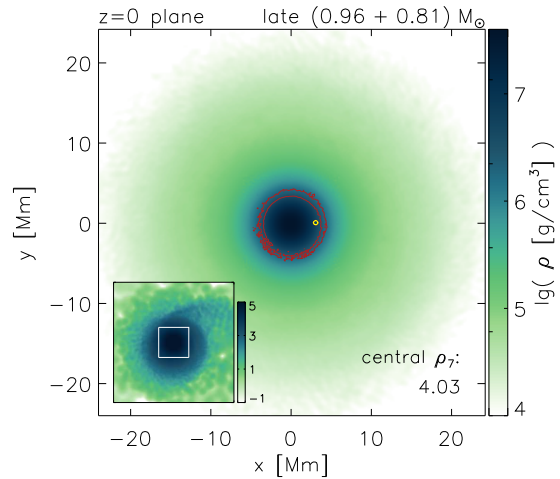}
\caption{Density in the orbital plane at the beginning of the detonation
simulation for 0p9-0p8 with \textsc{castro}. The yellow and red contours, at $7\times10^8\unit{K}$, shows the
location of the manually planted detonator in the merged object, and hot
regions at the interface with the surrounding disk. The inset shows a larger
region, with a white square indicating the boundaries of the main plot.}
\label{fig:initrho2}
\end{figure}

Figure~\ref{fig:initrho2} shows a snapshot of simulation 0p9-0p8c at the time when the detonation is set off.  There is hot material on the interface between the central post-merger object and the disk, indicated by red contours in the plot, but the temperatures are in general too low ($< 10^9\unit{K}$) to set off an autonomous detonation. The detonation is forcibly initiated by means of a spherical detonator (indicated by the yellow circle in Figure \ref{fig:initrho2}) of radius $300\unit{km}$, central temperature $1.6\times10^9\unit{K}$, and outer velocity $8\times10^8\cms$ near the surface of the merged object (on the $y=z=0$ line at $x_8=3$, where the density is about $5\times10^6\gcc$, see Figure~\ref{fig:initrho2}). With the exploding star being surrounded by a disk in every direction in the orbital plane, the ashes are accelerated most effectively along the polar axis.

\begin{figure}[ht]
\centering
\includegraphics[width=0.40\textwidth]{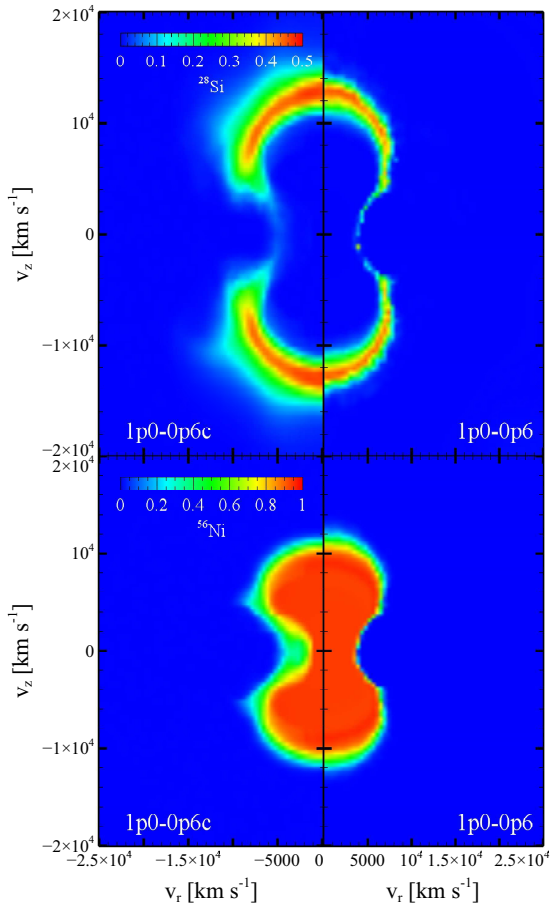}
\caption{Comparison of the azimuthally averaged distributions of \silicon[28] and \nickel[56] during the homologous expansion phase for simulations 1p0-0p6c (left sides of panels) and 1p0-0p6 (right sides). The two methods, \textsc{castro} and \textsc{snsph}, produce very similar results, despite the \textsc{castro} simulations featuring a surface detonation and the \textsc{snsph} simulations featuring centrally ignited remnants.}
\label{fig:homology}
\end{figure}

Simulation 1p0-0p6c was similarly detonated by means of a spherical detonator near the surface of the remnant core. The nucleosynthetic yield for the \textsc{castro} simulations are similar to that of the \textsc{snsph} result, with 0p9-0p8c resulting in $\approx0.72$\msol of \nickel[56] and $\approx0.02$\msol of stable \iron[54] and 1p0-0p6c resulting in $\approx0.80$\msol of \nickel[56] and $\approx0.03$\msol of stable \iron[54]. Comparing the ejecta velocities and isotopic distribution as in Figure \ref{fig:homology} for 1p0-0p6(c), we find very close agreement between the \textsc{snsph} and \textsc{castro} detonation results, despite the \textsc{snsph} detonations being centrally ignited. The nucleosynthetic yields for our \textsc{castro} simulations are shown in Table \ref{table:castroyields}.

\begin{table}[ht]
\caption{Isotope yields and kinetic energy for our \textsc{castro} simulations.}
\centering
\begin{tabular}{r | r r}
\hline\hline
 & 1p0-0p6c & 0p9-0p8c\\
\hline
\helium[4] & $6.33\times10^{-3}$ & $5.97\times10^{-3}$\\
\carbon[12] & $0.293$ & $0.274$ \\
\oxygen[16] & $0.373$ & $0.389$ \\
\neon[20] & $9.91\times10^{-4}$ & $1.06\times10^{-3}$ \\
\magnesium[24] & $1.55\times10^{-2}$ & $1.75\times10^{-2}$ \\
\silicon[28] & $0.100$ & $0.191$ \\
\el{S}{32} & $5.71\times10^{-2}$ & $0.106$ \\
\el{Ar}{36} & $8.98\times10^{-3}$ & $1.42\times10^{-2}$ \\
\el{Ca}{40} & $8.35\times10^{-3}$ & $1.19\times10^{-2}$ \\
\el{Ti}{44} & $1.14\times10^{-5}$ & $1.34\times10^{-5}$ \\
\el{Cr}{48} & $2.53\times10^{-4}$ & $3.21\times10^{-4}$ \\
\el{Fe}{52} & $5.76\times10^{-3}$ & $7.33\times10^{-3}$ \\
\el{Fe}{54} & $3.17\times10^{-2}$ & $2.46\times10^{-2}$ \\
\el{Ni}{56} & $0.799$ & $0.718$ \\
\hline
KE [erg] & $1.42\times10^{51}$ & $1.47\times10^{51}$\\
$28<A<40$ & $0.175$ & $0.324$ \\
$A\geq44$ & $0.837$ & $0.751$ \\
\hline
\end{tabular}
\label{table:castroyields}
\end{table}

\subsection{\nickel[56] Production and Viscous Evolution}

\begin{figure}[ht]
\centering
\includegraphics[width=0.40\textwidth]{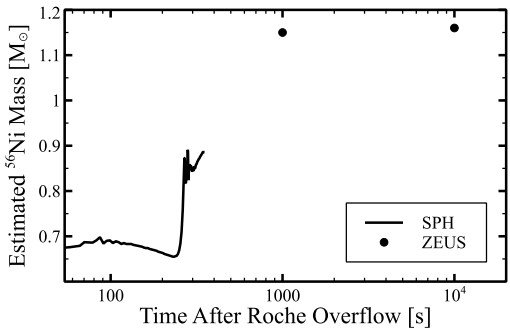}
\caption{The estimated \nickel[56] production for various times throughout the merger and viscous evolution of simulation 1p0-1p0(z). After the companion is completely disrupted, there is a short period of pulsations in the primary core while it adjusts to the new gravitational potential. This is reflected in the uncertainty of the \nickel[56] yield at $\approx300$ s.}
\label{fig:nickel}
\end{figure}

We find that the \nickel[56] production in post-merger detonations is greater than one in which the WD primary detonates during the merger itself \citep{Moll+2013}.  This is because the coalescence with the disrupted secondary WD leads to a compression of the WD primary, increasing its central density. 
We estimate the nucleosynthetic yields from these results using the \textsc{kepler}-compiled lookup table.  Figure \ref{fig:nickel} shows that this compression occurs quite rapidly a few 100 seconds after Roche overflow, and increases the estimated \nickel[56] yield by about 30\% for simulation 1p0-1p0. In \cite{Moll+2013}, a peri-merger detonation of simulation 0p9-0p8 in \textsc{castro} yields 0.58\msol of \nickel[56], while the post-merger detonation (included here as 0p9-0p8c) yields 0.72\msol, a 24\% increase.

Following the merger, the structure of the remnant will be further modified as the disk viscously evolves.  We also wish to examine the effects of the viscous evolution in order to explore how the production of \nickel[56] and \silicon[28]  changes as a function of the time between the merger and subsequent detonation. \cite{Schwab2012} recently simulated this phase using the staggered-mesh code \textsc{zeus-mp2} \citep{Hayes2006} (hereafter called simply \textsc{zeus}) in 2D. Here we employ the same methods following the procedures laid out in \cite{Schwab2012}. This code utilizes the Helmholtz EOS and approximates magnetic stresses via a shear stress term,
\be
T_{ij}=\rho\nu(\partial_i v_j + \partial_j v_i),
\ee
where $\nu$ is the dynamic viscosity coefficient;
\be
\nu = 3\times10^{-2}\frac{c^2_s}{\Omega_k},
\ee
with $c_s$ denoting the local sound speed and $\Omega_k$ the Keplerian angular velocity.  As in \cite{Schwab2012}, only the azimuthal ($T_{r\phi}$ and $T_{\theta\phi}$) components in the stress tensor were retained.

\begin{figure}[ht]
\centering
\includegraphics[width=0.40\textwidth]{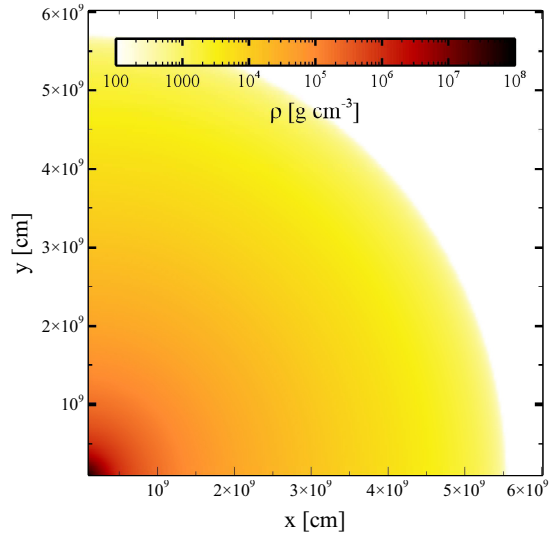}
\caption{The density distribution in the \textit{x-y} plane of simulation 1p0-1p0 after $10^4$ s. Viscous evolution has spherized the system, erasing the disk.}
\label{fig:1p0-1p0evo}
\end{figure}

The merged end-state of many of our \textsc{snsph} simulations are initialized in \textsc{zeus} on a spherical polar grid, with logarithmic grid spacing in the radial direction. After many viscous times ($\approx10^4$ s), the core density for each simulation rises precipitously, with \eg 1p0-1p0z (the ``z'' in this case referring to the \textsc{zues} counterpart to \textsc{snsph} simulation 1p0-1p0) rising from $7.09\times10^7$ g cm$^{-3}$ to $2.41\times10^8$ g cm$^{-3}$. The disks in these simulations evolve to a more spherical shape, as shown in Figure \ref{fig:1p0-1p0evo}. This more closely resembles the tamped configurations in the 1D models of \cite{Khokhlov1993} and \cite{Hoeflich1996}.

We estimate the nucleosynthetic yields from these results by interpolating the \textsc{kepler} table results to the final density profiles of the \textsc{zeus} simulations. For 1p0-1p0z, we find a $\approx30\%$ increase in the \nickel[56] yield after $10^4$ s, and a commensurate drop in \silicon[28] production by more than half (see table \ref{table:yields}).  As Figure \ref{fig:nickel} shows, depending on the time to detonation, the \nickel[56] yield can vary by as much as $80\%$. Simulation 1p2-1p0z also exhibits the expected decrease in \silicon[28], however the increase in \nickel[56] is not quite as substantial, with more material forming stable \iron[54], as indicated in Table \ref{table:yields}, which lists the \nickel[56], \iron[54], and \silicon[28] yields from each of our simulations.

\begin{table}[ht]
\caption{Isotope yields and kinetic energy for each of our simulations for which we initiated detonations, in addition to estimated yields (italics) from viscously evolved remnants. Simulations labeled with a ``c'' are \textsc{castro} results, while those labeled with a ``z'' are results from \textsc{zeus}.}
\centering
\begin{tabular}{c | c c c | c c c}
\hline\hline
\# & \nickel[56] & \iron[54] & \silicon[28] & KE [erg]\\
\hline
0p9-0p6   & 0.69 & 0.02 & 0.16 & 1.29$\times10^{51}$ \\
0p9-0p8   & 0.66 & 0.02 & 0.22 & 1.35$\times10^{51}$ \\
0p9-0p8c & 0.72 & 0.02 & 0.19 & 1.47$\times10^{51}$\\
0p9-0p8z & \textit{0.88} & \textit{0.03} & \textit{0.12}  \\
1p0-0p6   & 0.86 & 0.03 & 0.11 & 1.38$\times10^{51}$\\
1p0-0p6c & 0.80 & 0.03 & 0.10 & 1.42$\times10^{51}$\\
1p0-0p6z & \textit{0.91} & \textit{0.03} & \textit{0.10}  \\
1p0-1p0   & 0.89 & 0.03 & 0.26 & 1.65$\times10^{51}$ \\
1p0-1p0z & \textit{1.16} & \textit{0.06} & \textit{0.09}  \\
1p2-0p6   & 1.09 & 0.07 & 0.05 & 1.54$\times10^{51}$ \\
1p2-1p0   & 1.23 & 0.07 & 0.17 & 1.82$\times10^{51}$ \\
1p2-1p0z & \textit{1.35} & \textit{0.10} & \textit{0.06}  \\
1p0-0p4   & 0.70 & 0.02 & 0.12 & 1.27$\times10^{51}$ \\
\hline
\end{tabular}
\label{table:yields}
\end{table}

\section{Radiation-Transport}

In order to synthesize light curves and spectra for our detonation models, we use the radiative transfer code \textsc{~sedona} \citep{Sedona}. \textsc{sedona} uses a Monte Carlo technique wherein photon packets are emitted in the SN ejecta envelope which then scatter and absorb throughout a homologously expanding grid. The grid data is interpolated from the \textsc{snsph} results; given the near perfect axial symmetry of these models, the models were azimuthally averaged and the radiation transport calculations run in 2D.  The source geometry for the photon packet flux is first determined by energy deposition from the radioactive decay of \nickel[56] and \cobalt[56] and by any shock-heated gas. These packets then propagate throughout the domain where scatterings and absorptions are calculated from the opacities and emissivities of each cell they encounter. Temperatures for each cell are calculated in an iterative way by fixing the thermal emission rate to the calculated rates of photon packet absorption plus any surplus energy from radioactive decay. All photon packets that escape the system along certain lines of sight are used to construct synthetic light curves and spectra.

\subsection{Light Curves \& Spectra}

\begin{figure*}[ht]
\centering
\includegraphics[width=0.7\textwidth]{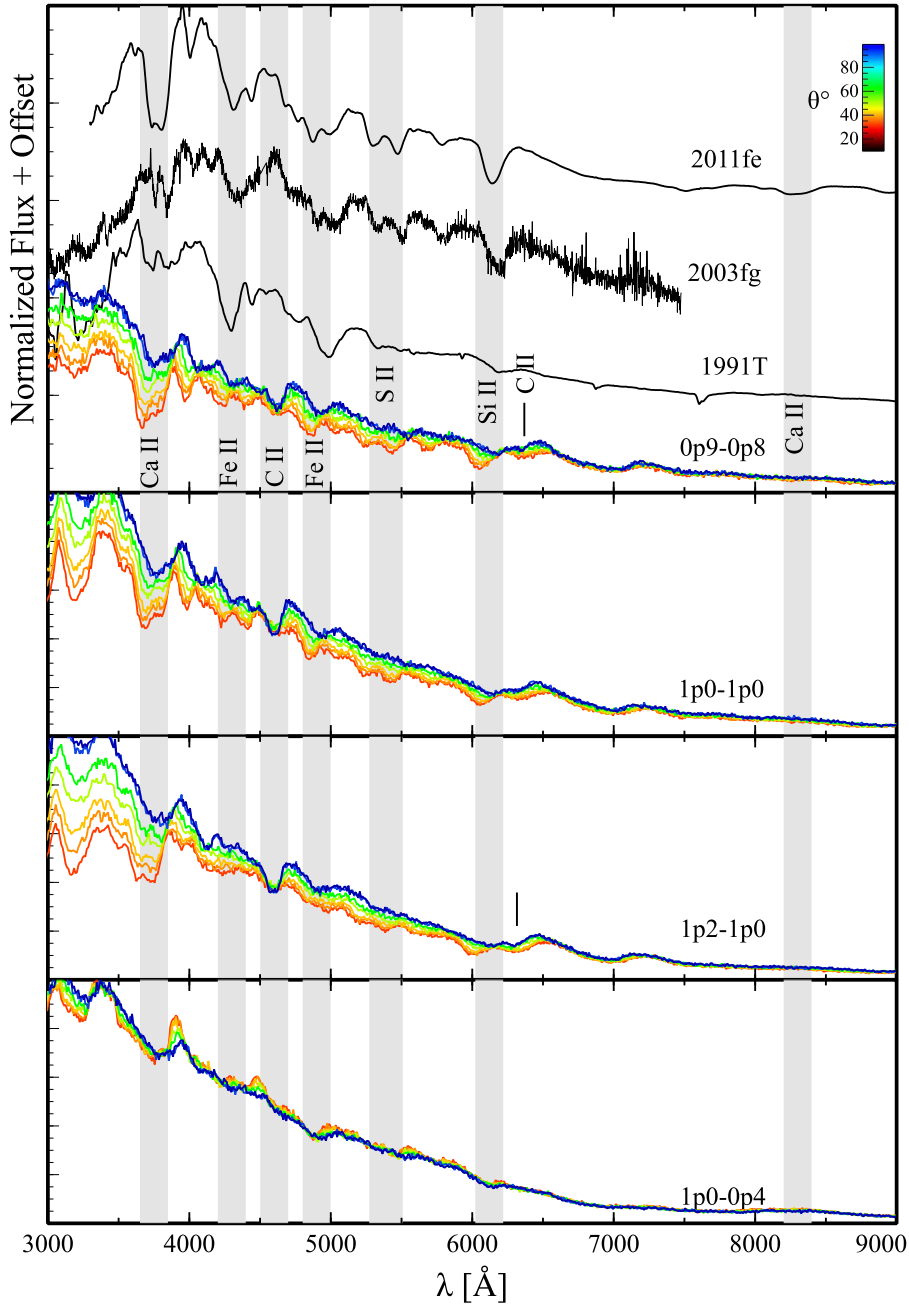}
\caption{Near maximum-light synthetic pectra from simulations 0p9-0p8, 1p0-1p0, 1p2-1p0, and 1p0-0p4 at several viewing angles from $14.8\degree<\theta<91.9\degree$. High values of $\theta$ correspond to lines of sight through the accretion disk. Several prominent absorption features are indicated with gray bars, and comparison spectra for a standard SNIa (SN2011fe; \cite{SN2011fe}), a peculiar SNIa (SN1991T; \cite{SN1991T}) and a superluminous SNIa (SN2003fg; \cite{SN2003fg}) are plotted with offsets.}
\label{fig:spectra}
\end{figure*}

Figure~\ref{fig:spectra} plots the near maximum light synthetic spectra  for a sample of the models.  The color coding shows the orientation which varies from a viewing angle along the pole ($\theta \approx 0\degree$) to one one along the equator  ($\theta \approx 90\degree$).  The model ejecta exhibit a nearly perfect top/bottom reflective symmetry, and so the spectra as seen from viewing angles for $90\degree < \theta < 180\degree$ are essentially identical to those with $\theta < 90\degree$.  

Several strong orientation effects can be seen in the synthetic spectra of models 0p9-0p8, 1p0-1p0, and 1p2-1p0. First, the overall luminosity is higher from the equatorial views, a result of the larger projected surface area of the ejecta when viewed from these angles (see Figure~\ref{fig:isotopes}).    Second, the Doppler shifts of most absorption features are lower for equatorial views, a result of the deceleration of the ejecta by the surrounding disk.  For model 0p9-0p8, the velocity (as measured from the minimum of the SiII line with rest wavelength 6355~\AA) is $v \approx 15,000~\kms$ for  $\theta = 0\degree$ but only 9600~\kms for $\theta = 0\degree$.  Third, the IME absorption features are generally weaker from the equatorial views.  This is also due to the deceleration of the ejecta, which narrowed the velocity range of IMEs above the  photosphere.  Fourth, the continuum is significantly bluer and the ultraviolet (UV) flux is much higher from viewing angles near $\theta \approx 0\degree$. This is presumably due to the reduced line blanketing from iron group elements for this orientation.  The disk interaction slows the \nickel[56] ejecta to  $\sim 5000~\kms$ in the equatorial regions, which is well below the photosphere at these epochs.  In the absence of fast-moving iron, the line blanketing is weak and the light emerges in large part at blue and UV wavelengths.

The spectra of  model 0p9-0p8 viewed near the pole ($\theta<45\degree$) are fairly similar to standard SNeIa (represented here by SN2011fe), though the absorption features are slightly higher velocity than normal. However, for the lines of sight nearer the equator ($\theta>45\degree$),  the spectra  lack strong SiII and SII features, and more closely resemble that of the peculiar SN~Ia 1991T \citep{SN1991T}.  However, unlike SN~1991T, strong CII absorption features are seen in the model spectra near $6300$~\AA\ and $\approx4700$ \AA.  This reflects the large mass of unburned carbon/oxygen in the swept up  disk.   The spectra of models 1p0-1p0 and 1p2-1p0 show similar trends, but have weaker IME absorption features from the polar angles.

The spectra for model 1p0-0p4 are peculiar and nearly featureless at all viewing angles. This weak orientation dependence is the result of the 0.4\msol companion star forming a more spherical configuration than in other simulations (see Table \ref{table:heights}). While iron absorption appears somewhat faster than standard, the SiII line, though shallow, has a fairly normal velocity. CaII and SII lines are suppressed.  No helium lines are visible in the the synthetic spectra, despite the large mass of unburned helium remaining in the outer layers of the ejecta, however this may be the result of the neglect of non-thermal excitation in these LTE radiative transfer calculations \citep{Lucy_1991}. 

\begin{figure*}[ht]
\centering
\begin{tabular}{rl}
\includegraphics[width=.4\linewidth]{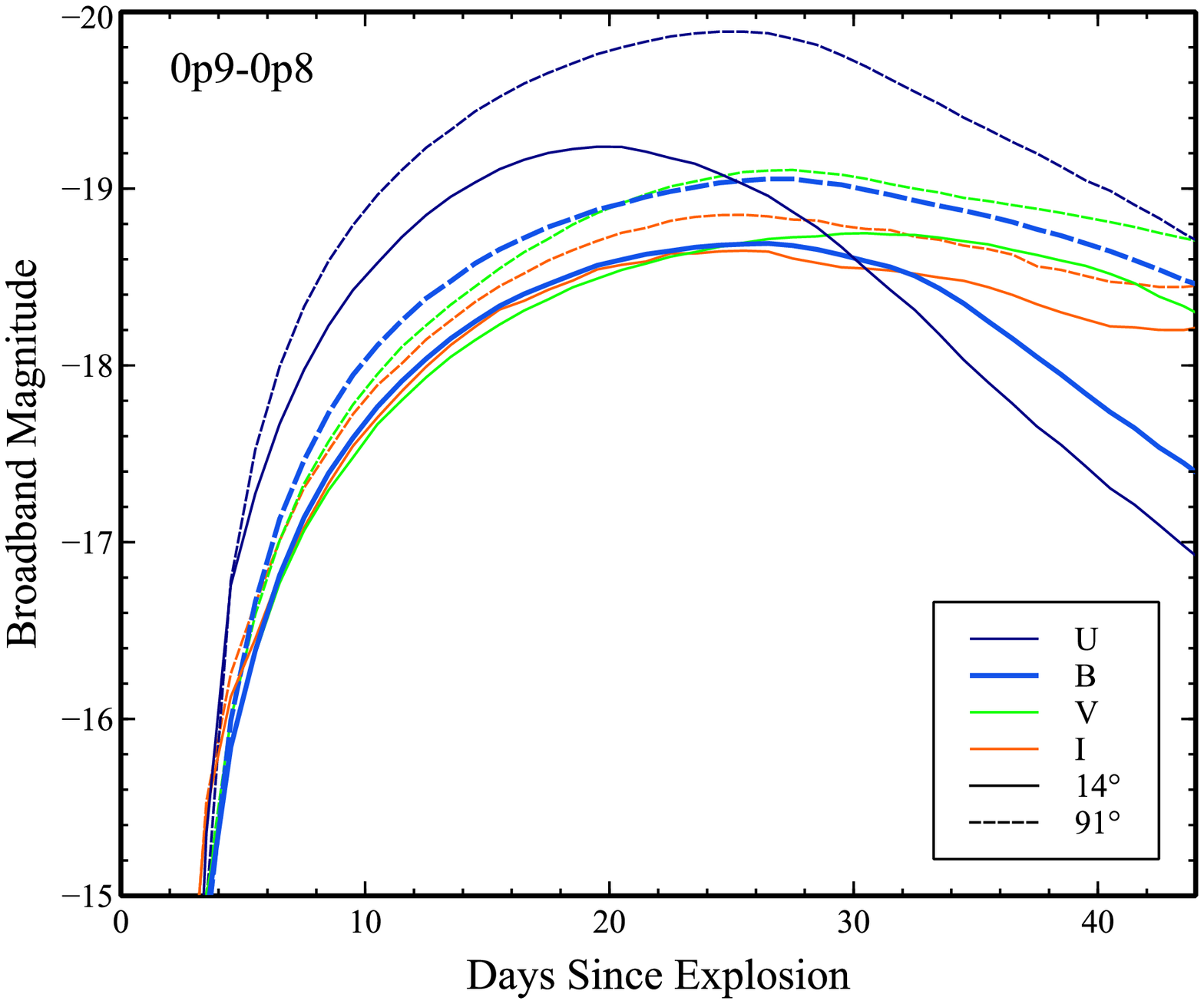} &
\includegraphics[width=.4\linewidth]{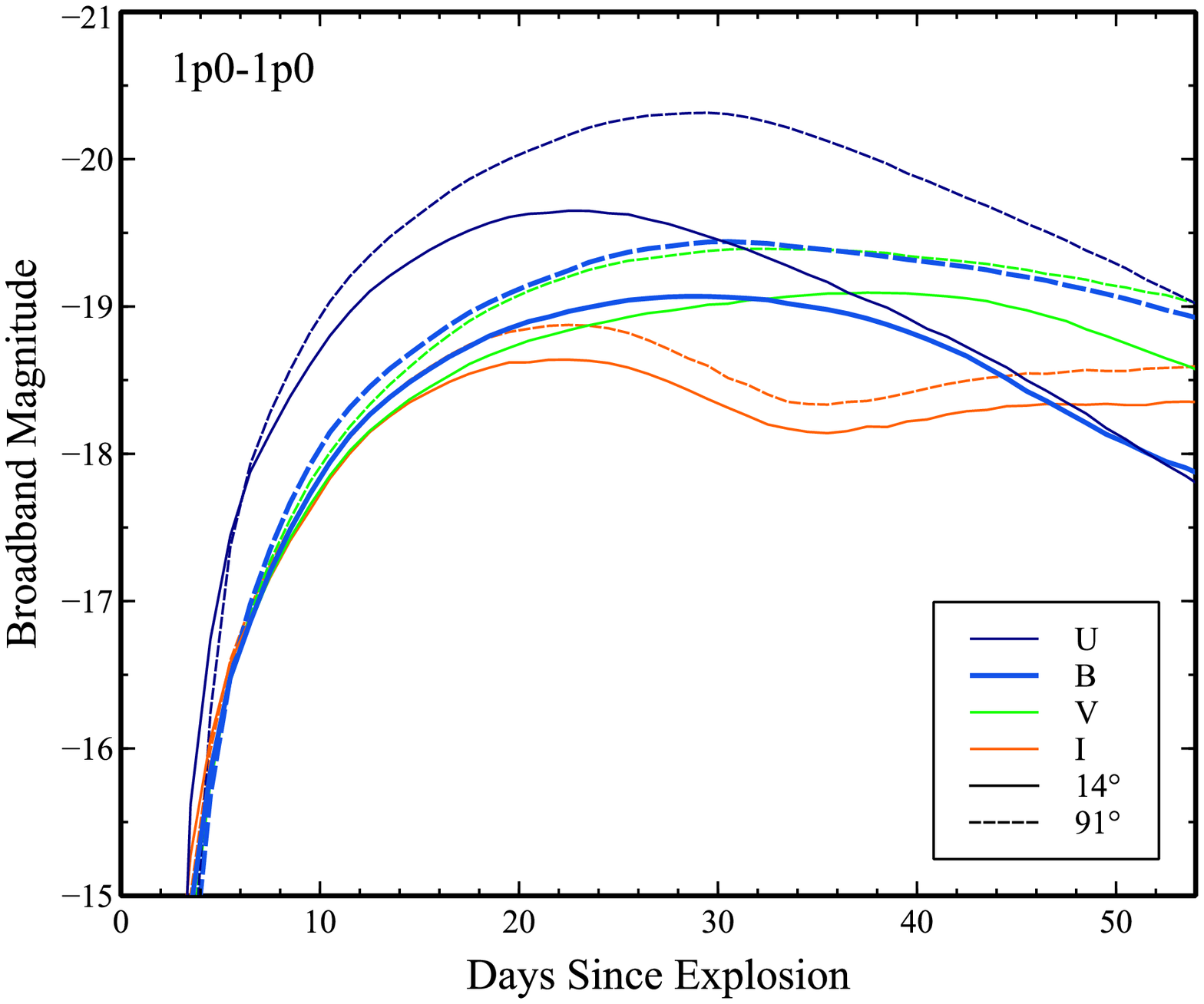}\\
\includegraphics[width=.4\linewidth]{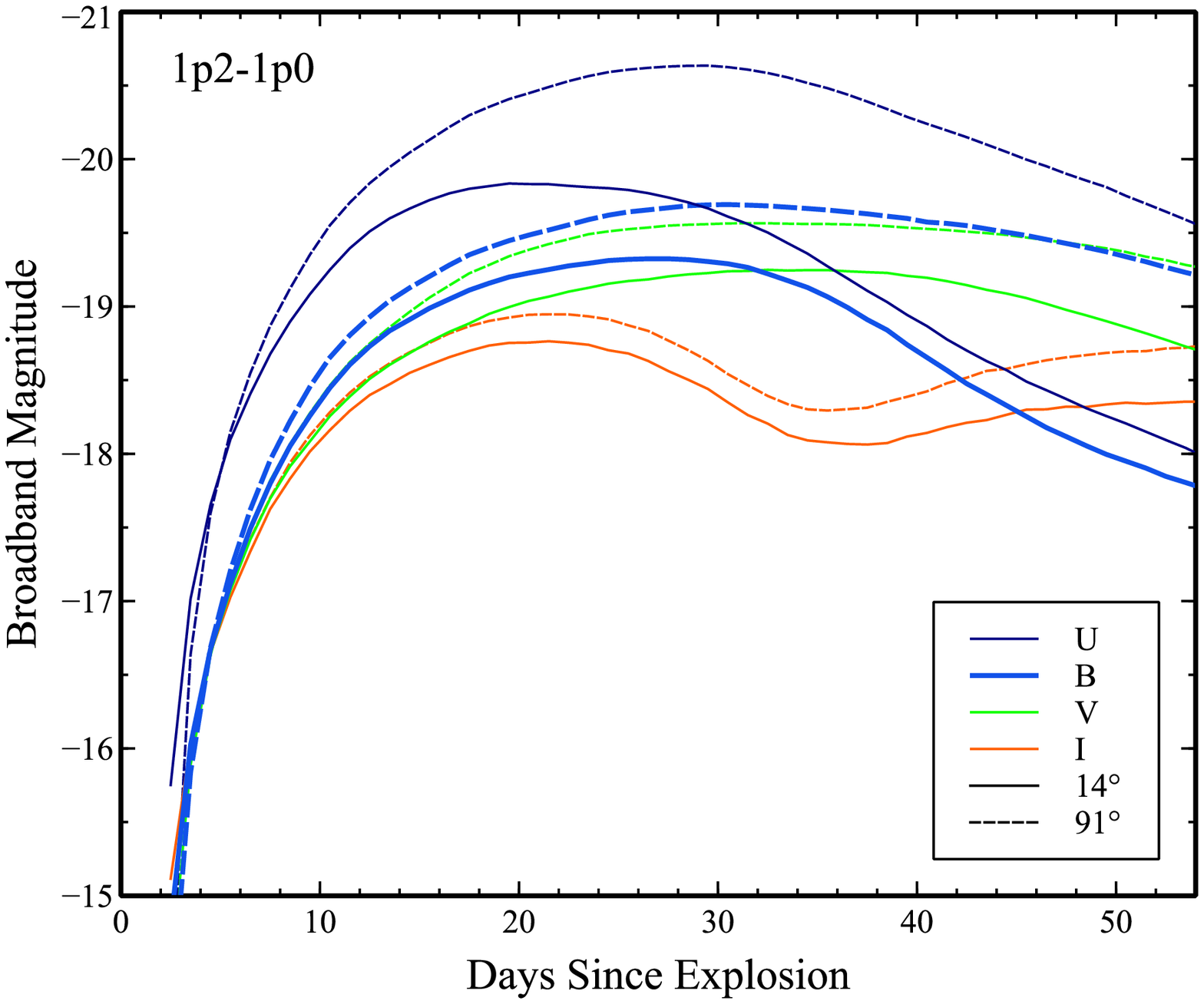} &
\includegraphics[width=.4\linewidth]{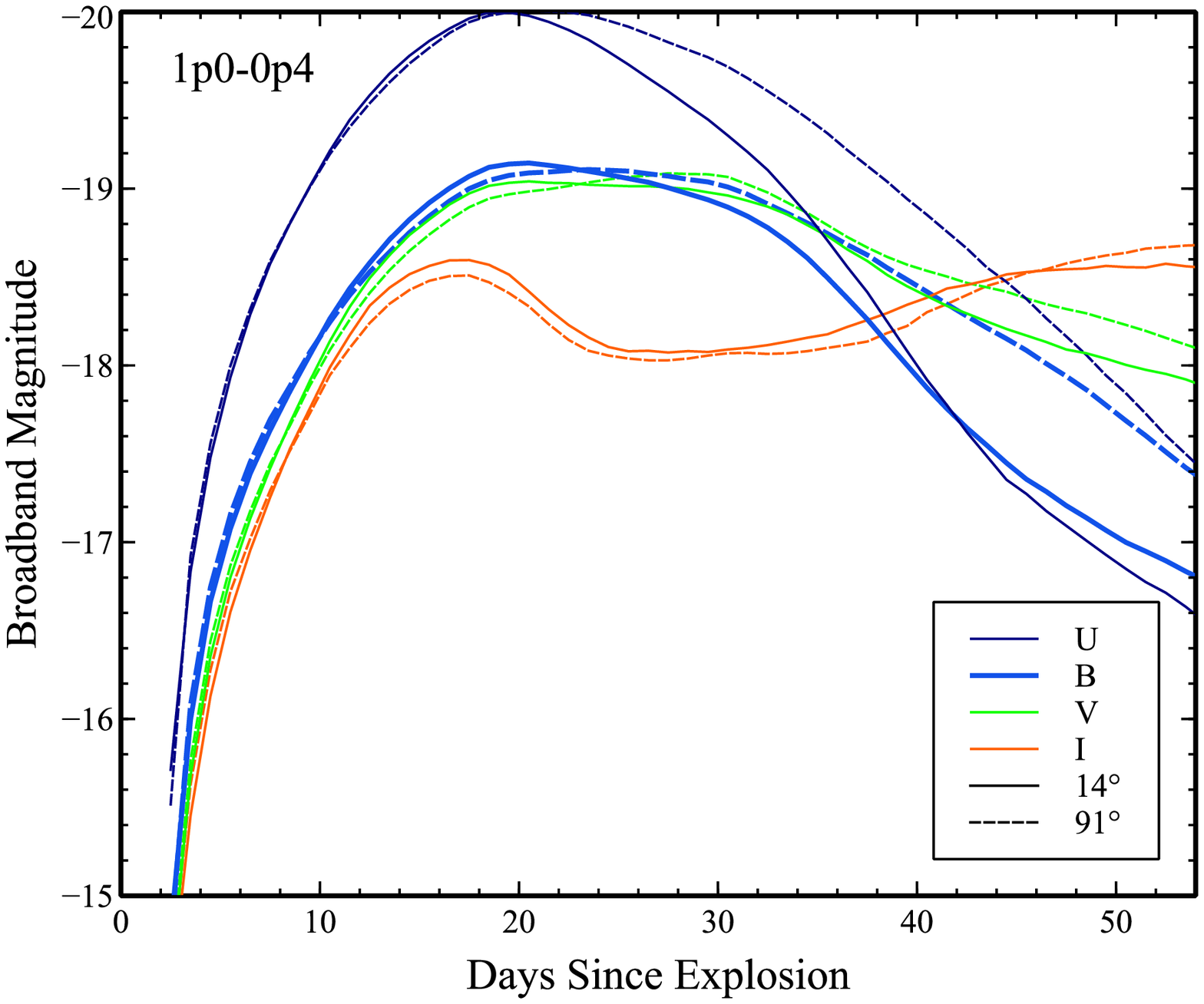}
\end{tabular}
\caption{Lightcurves in the U, B, V, and I bands from simulations 0p9-0p8, 1p0-1p0, 1p2-1p0, and 1p0-0p4 at $\theta=14\degree$ (solid lines) and $\theta=91\degree$ (dashed lines).}
\label{fig:lightcurves}
\end{figure*}

Figure \ref{fig:lightcurves} plots the synthetic broadband light curves for several of our simulations as viewed from either a polar or an equatorial angle.  Table \ref{table:mags} also lists the B-band magnitudes at peak and the decline rate $\Delta$m$_{15}(B)$ of the light curves at these two angles.  Both the light curves rise times and decline rates are slow compared to those of normal SN~Ia.  This reflects the relatively long effective diffusion time in these models,  which have a larger total ejecta mass and a larger fraction of unburned material than standard \Mch\ explosion models.  The post-maximum B-band light curve is also strongly influenced by line-blanketing, which affects the decline rate by progressively redistributing flux to longer wavelengths \citep[e.g.,][]{Kasen_Woosley_2007}.  From the equatorial viewing angles, the low velocity of the iron group elements results in reduced line blanketing, and hence a slower B-band decline rates from these orientations.

\begin{figure}[ht]
\centering
\includegraphics[width=0.40\textwidth]{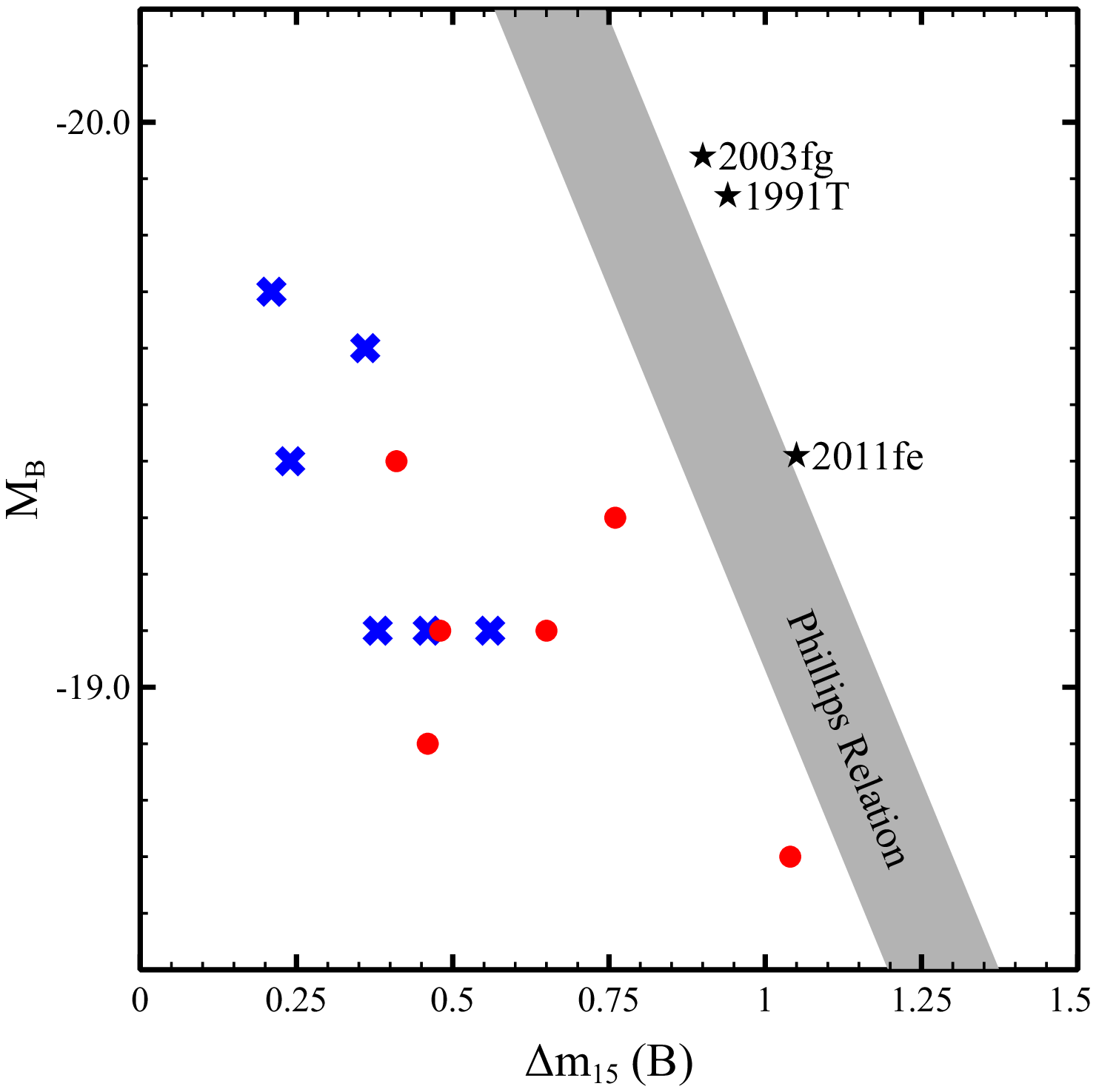}
\caption{$\Delta$m15 vs.\ peak B magnitude for each of the CO-CO simulations. Blue crosses correspond to a viewing angle of $\theta=91\degree$ while red circles correspond to $\theta=14\degree$. A linear approximation to the Phillips Relation \citep{Phillips1993,HighEAstro2011} is plotted for comparison.}
\label{fig:phillips}
\end{figure}

Figure \ref{fig:phillips} plots the width-luminosity relation (i.e., peak B-band magnitude vs. the decline rate $\Delta$m$_{15}(B)$) 
of the models.   The peak B-band magnitude of the models vary by as much as 0.4~mag depending
on the viewing angle.   From angles near $\theta = 0\degree$, the
 more massive models, in particular 1p2-1p0, predict a peak B-band magnitude approaching
the brightness of the observed super-\Mch\ events.  However, the B-band decline rate of the models is 
smaller than that observed for specific cases such as SN~2003fg.  
As a whole, the decline rate of the models is typically too slow, given 
their peak brightnesses, when compared to the observed Phillips relation. 

\begin{table}[ht]
\caption{Peak B-band magnitude and $\Delta$m$_{15}$ for each of our simulations for which we constructed light curves and spectra at $\theta=14\degree$/$\theta=91\degree$. The three SN whose spectra are shown in Figure \ref{fig:spectra} are also listed for comparison.}
\centering
\begin{tabular}{c | c c c | c c c}
\hline\hline
\# & Bmag & $\Delta$m$_{15}$(B)\\
\hline
0p9-0p6   & -18.9/-19.1 & 0.46/0.38\\
0p9-0p8   & -18.7/-19.1 & 1.04/0.46\\
1p0-1p0   & -19.1/-19.4 & 0.48/0.24\\
1p2-0p6   & -19.4/-19.6 & 0.41/0.36\\
1p2-1p0   & -19.3/-19.7 & 0.76/0.21\\
1p0-0p4   & -19.1/-19.1 & 0.65/0.56\\
\hline
1991T  & -19.87 & 0.94\\
2003fg & -19.94 & 0.9\\
2011fe & -19.41 & 1.05\\
\hline
\end{tabular}
\label{table:mags}
\end{table}

\section{Discussion}

We have brought to bear a number of computational tools to study the effects of various remnant configurations from double-degenerate mergers on a SN that might explode within them. First, we have shown that a generic range of disk configurations can result from CO-CO mergers. We find that the disk scale height increases with the mass ratio ($q$) of the binary progenitor system. This scale height plays a critical role in the line of sight variability of the resulting SN spectra. We also find that mergers with low-mass helium companions can still occupy a portion of parameter space where binary mergers are unstable. Moreover, such high-$q$ mergers form much more spherical envelopes, all but erasing the variability  of the observables with the line of sight. 

By comparing the results of centrally ignited models to surface detonators, we have found that the location of ignition in
post-merger detonation has little affect on the SN structure as compared to the presence of an accretion disk. Both our grid-based and particle-based detonation models produce very similar radioactive yields as well, with the grid-based detonations featuring only slightly greater kinetic energies. This small disparity may only be the result of the different methods for igniting the detonation as opposed to the location of the detonator or the differences between particle-based and grid-based hydrodynamics codes.

Our synthetic light curves and spectra of most models posses strong viewing angle dependences as a result
of the ejecta asymmetry imparted by the interaction with the CO disk.  
The one exception is the He + CO WD merger 1p0-0p4, where the remnant disk had a  large scale height due to the high-$q$ value.    In general, observations along the equatorial region ($\theta \approx 90\degree$) feature slower line absorptions, and brighter SNe with wider light curves. The sign of this trend conform to the Phillips Relation, but all of our models for which we have constructed light curves are much longer-lasting (have much smaller $\Delta$m$_{15}$ values) than  standard SN~Ia.   
The maximum light spectra were peculiar from some viewing angles, showing weak IME absorptions and relatively strong
CII absorptions.

 Tamped WD explosions such as those studied here have often been invoked to explain  the class of 
 very luminous, super-\Mch\ SN~Ia. In this context, our models have certain interesting properties:  the surrounding CO disk decelerates the ejecta while remaining unburned, leading to
relatively lower IME absorption velocities, and strong CII lines when the event is viewed from the equatorial  regions.  In addition, the light curves are also brighter from these viewing angles, by as much as 40\%, due to the
larger projected surface area of the hourglass shaped ejecta.  
While these trends all have the right sense to explain the super-\Mch\ events,  
the particular models considered here do not succeed in quantitatively reproducing all of the observed properties.  In particular,
even the most massive model we consider, 1p2-1p0, is slightly dimmer ($\sim 0.2$~mag)  in the B-band than SN~2003fg.  In addition, the B-band decline is much slower than observed and the colors are significantly bluer.  
Despite these discrepancies, it is notable that the ejecta asymmetries that  arise in post-merger
explosions can lead to both enhanced luminosity and lower ejecta velocities from
certain viewing angles.  The high \nickel[56] masses inferred for some of the super-\Mch\ events 
($M_{\rm ni} \approx 1.4-1.8~\Msun$) may therefore be an overestimate of what is truly required to explain
the observed brightness.  

We found that for post-merger detonations have an enhanced \nickel[56] production compared to peri-merger detonations, due to compression of the primary WD in the merged remnant.  Our viscous evolution simulation demonstrate that additional compression occurs in the subsequent hours.  Thus, explosion occurring with longer delay times after the merger event can result in $\sim 80\%$ greater production of \nickel[56]. However, longer delay times will also result in much more spherical explosions. If the system were to explode in this state, or at some subsequent phase, the observables would be distinct from those considered here. In further studies, we will consider the impact of these more evolved systems on the likelihood of detonations and their ejecta structure and spectra.

As the precise timing between WD mergers and subsequent SN~Ia is still an open question, explorations of all possible outcomes are important for both populating and constraining the menagerie of transients.  Some portion of parameter space in Nature may produce SNe like those explored here.  With the recent deluge of new and unusual transient observations from wide-field surveys, models like these may prove valuable for their classification.

\section*{Acknowledgments}
This research has been supported by the DOE HEP Program under contract
DE-SC0010676; the National Science Foundation (AST 0909129 and AST-1109896) and the
NASA Theory Program (NNX09AK36G).  DK is supported by a Department of Energy Office
of Nuclear Physics Early Career Award (DE-SC0008067). Rainer Moll acknowledges support by the Alexander von Humboldt
Foundation through the Feodor Lynen Research Fellowship program. JS is supported by an NSF Graduate Research
Fellowship.
We thank John Bell and Ann
Almgren for their major roles in developing the CASTRO code.  This
research used resources of the National Energy Research Scientific
Computing Center, which is supported by the Office of Science of the
U.S. Department of Energy under Contract No. DE-AC02-05CH11231.  This
research used resources of the Oak Ridge Leadership Computing Facility
at the Oak Ridge National Laboratory, which is supported by the Office
of Science of the U.S. Department of Energy under Contract
No. DE-AC05-00OR22725.
We are grateful for computer time supplied by the Advanced Computing Center at Arizona State University, and to Frank Timmes for technical support and for insightful discussions during the construction of our nuclear network. 

\bibliographystyle{apj}

\end{document}

%% file: nuclides.tex
\newcommand{\nuclei}[2]{\ensuremath{\mathrm{^{#1}#2}}}

\newcommand{\helium}[1][4]{\nuclei{#1}{He}}

\newcommand{\carbon}[1][12]{\nuclei{#1}{C}}

\newcommand{\oxygen}[1][16]{\nuclei{#1}{O}}

\newcommand{\neon}[1][20]{\nuclei{#1}{Ne}}

\newcommand{\magnesium}[1][24]{\nuclei{#1}{Mg}}

\newcommand{\silicon}[1][28]{\nuclei{#1}{Si}}

\newcommand{\iron}[1][26]{\nuclei{#1}{Fe}}
\newcommand{\cobalt}[1][59]{\nuclei{#1}{Co}}
\newcommand{\nickel}[1][58]{\nuclei{#1}{Ni}}